\magnification=\magstephalf
\parskip=4pt
\baselineskip=12pt plus 1pt minus 1pt

\centerline{
{\bf From the Big Bang to the Multiverse: Translations 
in Space and Time}}
\centerline{by David H. Weinberg, Ohio State University, 
Department of Astronomy} 

\bigskip\noindent
This essay appears, in slightly modified form and with a much larger
and more beautiful set of accompanying images, in
{\it Josiah McElheny: A Prism}, edited by Louise Neri and Josiah McElheny,
Skira/Rizzoli Books, New York, 2010.

\bigskip
\noindent
\centerline{\bf An End to Modernity}
\medskip

  The Big Bang is an extraordinarily difficult subject for visual
representation.  When Josiah McElheny and I met for the first time,
in September 2004,
we began by examining his sketch for the piece that
would become, over the following year,
{\it An End to Modernity.}  In this sketch,
inspired by the spectacular Lobmeyr chandeliers that hang
in the Metropolitan Opera House, 
a central sphere supports a starburst of metal rods, which in turn
support clusters of glass pieces and lamps.
This seemed a perfect depiction of the popular
conception of the Big Bang, a tremendous explosion that flings
fragments of matter in all directions from a central point.
But the Big Bang is {\it not} an explosion of material into space;
it is the origin of space and time itself, initiating an expansion that
occurs everywhere and has no center.  How could any static
sculpture, no matter how intricate, depict {\it that}?

  During the ensuing three-hour conversation, the solution
emerged: retain the basic structure of the sketch, but change
the {\it interpretation} of that structure by using a spatial dimension
to represent time.  
The center of the sculpture would then become the primordial cosmos and
the outer edge the present day, and the passage from
one to the other would trace the 14 billion year history of the
expanding universe.
With this space-to-time translation, {\it An End to Modernity} could 
incorporate many of the key features of cosmic evolution that
have been revealed by astronomical observations over the last four
decades, following prescriptions that Josiah and I worked out
over many months of meetings, emails, and phone calls.
In the finished work, each of the 
230 radiating rods emerges in a random
direction with a randomly selected length and terminates 
in a cluster of hand-formed glass disks and blown-glass globes,
representing a cluster of galaxies, or else in a single lamp, representing 
a quasar.  The rules that govern the form, size, and contents of the 
cluster, or the brightness of the  lamp, depend on the length 
of the rod, and thus on the cosmic epoch
that corresponds to its termination point.
The clusters are surrounded, we must imagine, by
clouds of invisible ``dark matter,'' subatomic particles whose 
identity remains unknown but whose existence is inferred from
their observed gravitational effects on the rotations and motions of galaxies.

  The central aluminum sphere that supports the cluster and lamp rods
also represents one of the crucial
concepts of modern cosmology: the Last Scattering Surface.
The early universe was so hot that hydrogen atoms were broken apart 
into their constituent protons and electrons.  The free electrons 
scattered light much like the water molecules of a dense fog, making 
the universe opaque.  When the expanding cosmos cooled to a
temperature of 3000$^\circ$C some 500,000 years
after the Big Bang, all free electrons disappeared into atoms,
and the universe became transparent.  The most distant source of
``light" that we can see is the opaque ``surface" nearly 14 billion
light years away.
The 1965 discovery
of the cosmic microwave background, the faded glow of this surface
now cooled to a mere three degrees above absolute zero, provided 
the key piece of evidence for the hot early universe postulated
by the Big Bang theory.

  In {\it An End to Modernity}, the Big Bang itself is a conceptual
point at the center of the sculpture, time zero, hidden by the aluminum
sphere just as the electron fog of the last scattering surface shields
the earliest epochs of the cosmos from our direct view.
From the central sphere outwards, the 
sculpture incorporates a logarithmic mapping between distance from
the center and the size of the universe.  Every $7.2"$ of distance 
corresponds to a factor of two in cosmic expansion, and the 6-foot 
span from the central sphere to the outer edge of the sculpture
represents a 1000-fold growth in each spatial dimension.  

  At the epoch of last scattering, the universe was filled with smoothly 
distributed hot gas, with no galaxies or stars or living creatures to 
admire them.  Over time, gravity amplified small fluctuations created 
in the first trillionth of a second of cosmic history into the galaxies 
and larger structures
that we observe today.  In {\it An End to Modernity}, the first galaxies
appear three feet out from the central sphere, 100 million years after the
Big Bang.  Like our own Milky Way, these early 
galaxies have the disk-like shapes
that are the generic final state of a cooling,
rotating cloud of gas.  Moving outward towards the present, three 
things change.  First, the galaxies get bigger, as they attract gas from
their surroundings and process it into stars.  Second, gravity pulls
the galaxies themselves into ever larger structures, galaxy clusters
and superclusters.  Finally, a new type of galaxy begins to appear,
rounded elliptical systems formed by chaotic collisions of the
rotationally ordered disks.  Elliptical galaxies, depicted
by glass spheres, reside mainly in the dense clumps where 
collisions are most common, while the extended, filamentary superclusters
are populated mainly by disks.

  The lamps that illuminate {\it An End to Modernity} represent quasars,
the brightest objects in the universe.  Quasars are powered by supermassive
black holes, up to ten billion times the mass of the Sun, which reside
at the centers of galaxies.  As gas falls into these black holes, it
collides with itself at speeds close to the speed of light, heats up,
and emits light, X-rays, and other forms of radiation.  When a supermassive
black hole is actively devouring gas, it can outshine the combined starlight
of its host galaxy by a factor of 1000.  The first quasars are faint
because there has not been time to build the most massive black holes.
The height of the quasar era is the period 2-4 billion years after the
Big Bang.  At later times, the population slowly fades because the 
galactic disturbances that feed gas to the black holes become less common,
so that most galaxies harbor only a dark remnant of their former glory.
In {\it An End to Modernity}, this history is encoded by the changing
frequency and brightness of the quasar lamps.

  Because light travels at a finite speed, astronomical telescopes function
as time machines: when we observe distant objects, we see them not as
they are today but as they were when they emitted their light.  This
fortunate feature of physics allows us to build an empirical
picture of the history of the universe from observations at the
present day.  {\it An End to Modernity} represents the principal
elements of this picture --- the Last Scattering Surface, the growth, 
transformation, and clustering of galaxies, 
and the rise and fall of the quasar population ---
in idealized but qualitatively accurate form.
From our earthbound vantage point, the history of the universe is
traced through a series of concentric shells, with the earliest
epochs seen at the greatest distances.  {\it An End to Modernity}
inverts this perspective, and it invites us to view the history
of the cosmos as though we stand outside of it.
However, if current astronomical inferences are
correct, the universe will expand forever, at an ever accelerating
rate, driven outwards by the repulsive gravitational effect
of ``dark energy'' that fills otherwise empty space.
Thus, wherever we stand to regard {\it An End to Modernity},
we are implicitly enveloped by the future.

\bigskip
\noindent
\centerline{\bf The Last Scattering Surface}
\medskip

{\it An End to Modernity} depicts, in idealized form, the entire history
of the cosmos.  {\it The Last Scattering Surface} zeroes in on two
special moments of that history: the present day, and the epoch
of recombination, when the formation of hydrogen atoms lifted
the free electron fog that had prevailed for the preceding
half million years.
With the space-to-time translation that underlies both 
{\it An End to Modernity} and {\it The Last Scattering Surface},
the epoch of recombination is represented by the central
aluminum sphere, the opaque barrier that conceals the
early universe within.  But the real last scattering surface
is not merely a screen; it glows with the heat of the primordial
cosmic fireball.  At the recombination temperature of $3000^\circ$ C,
this glow would have a dull orange color.
The subsequent expansion of the universe
has stretched the primordial light to millimeter wavelengths,
where it is no longer visible to our eyes but can be measured by 
microwave radio telescopes.
For more than 25 years after its initial discovery, steadily
improving observations showed that 
the cosmic microwave background is almost perfectly uniform
across the sky.  This uniformity tells us that the early
universe was full of smoothly distributed hot gas,
with no galaxies, stars, or planets, or living beings
to admire them.

In 1992, the {\it Cosmic Background Explorer} ({\it COBE}) satellite
made the first detection of intensity variations on the
last scattering surface, finding minute temperature differences of 
one thousandth of one percent from one point on the sky
to another.  These temperature differences are the imprint
of tiny variations in the underlying density of matter.
Over the last 14 billion years, gravity has
pulled matter into the denser regions,
a snowballing process that amplified the tiny
seeds present at recombination into the large scale cosmic
structure that we observe today.  The intensities of the
lamps that illuminate {\it The Last Scattering Surface} are
chosen based on the {\it COBE} image of the microwave sky --- there
is a one-to-one mapping between the locations of the lamps
and points on the true celestial sphere.  The clusters of glass
pieces at the outer edge represent present-day clusters of spiral 
and elliptical galaxies, just as they do in {\it An End to Modernity}.  

The structure observed in the cosmic microwave background 
is ``scale-invariant'': if we blur the {\it COBE} map,
smoothing over its small scale roughness, then the
level of intensity variation barely decreases.
This scale-invariance carries through to 
{\it The Last Scattering Surface}: if a lamp's neighbors are
bright, then it is more likely than average to be bright as well.
We see variations from one lamp to the next, but also patches
that are coherently bright or dim, and even, as we walk around
the perimeter, whole octants of the sphere that are ``hot''
or ``cold.''  The scale-invariance of primordial fluctuations,
processed subsequently by the gravity of radiation and matter,
is the reason that galaxies do not pepper the universe 
randomly but are instead drawn into an intricate filamentary web.

  A perfectly smooth universe could remain perfectly smooth, and 
perfectly boring, forever.  {\it The Last Scattering Surface} depicts the 
subtle imperfections of the primeval cosmic glow, and the rich, beautiful 
structures that emerged from them.

\bigskip
\noindent
\centerline{\bf The End of the Dark Ages}
\medskip

After the expanding cosmos cooled to $1000^\circ$ C, the background
radiation stretched to infrared wavelengths, and the universe was
devoid of visible light.  Where the initial concentration of atoms 
and dark matter was unusually high, gravity pulled matter together into 
dense, slowly spinning clumps.  Some 5-10 million years after the
Big Bang, the cores of the densest contracting gas clouds heated to the 
temperature needed for nuclear fusion to begin, and the first
stars in the universe were born.

{\it The End of the Dark Ages} zeroes in on the first one billion
years of cosmic history, when the diffuse glow of the Big Bang faded and the
first stars, galaxies, and quasars filled the universe with discrete
sources of visible light.  In structural terms, it is like the
inner regions of {\it An End to Modernity}, expanded in scale and
detail to allow closer inspection.  All of its galaxies are small,
since they have not yet had time to grow to larger size.
Nearly all of them have the disk-like form expected from the collapse
of spinning gas clouds, since they have not had time to fragment
completely into stars and merge with each other to create disordered
elliptical systems.  Lamps mark the first quasars, massive black holes
that glow as they swallow gas from their galactic hosts.
These early quasars outshine the surrounding galaxies, but they are
not as massive or as bright as the super-luminous systems that
will succeed them two billion years later.
The first galaxies and quasars are clustered on large scales
because they form at rare locations, the high 
mountain peaks of the primordial matter distribution that are
most susceptible to early collapse.
However, gravity has not yet arrayed the galaxies into filamentary
superstructures.

Astronomers define the end of the cosmic ``dark ages'' as the
time when ultraviolet radiation from hot stars and accreting
black holes penetrated into the depths of intergalactic space,
breaking the hydrogen atoms that resided there back into
protons and electrons.  This ``epoch of reionization'' --- the
inverse of the recombination that formed the last scattering
surface --- appears to have occurred $400-800$ million years
after the Big Bang.  However, even the deepest images from
{\it Hubble Space Telescope} and the largest ground-based 
telescopes reach only to the end of the reionization epoch.
The structure of {\it The End of the Dark Ages} is therefore
based mainly on theoretical modeling, computer calculations that
draw on what we know about atoms, dark matter, and ripples in
the cosmic microwave background to predict how the first stars
and galaxies formed.  {\it Hubble's} successor, the
{\it James Webb Space Telescope}, is designed to penetrate
further in space and deeper in time, to open the dark ages
to direct observation.

At the end of the dark ages, the universe was simpler than 
the one we inhabit today, the smooth skin of its youth
only lightly wrinkled by gravity.  But for the first time
in cosmic history, an alien being on a young world could have
looked up and seen a night sky much like our own, sparkling
with stars, and with galaxies beyond.

\bigskip
\noindent
\centerline{\bf Island Universe}
\medskip

Space is big.  This claim may seem intuitively obvious, but it can
be made precise: the observable universe 
contains roughly $10^{78}$ atoms and $10^{87}$ cosmic microwave
photons, numbers that are straightforward to calculate but
impossible to envision.  Despite the richness of structure traced by galaxies
and clusters of galaxies, on large scales the universe is remarkably
uniform.  We see the same kinds of structures whatever direction
we look in, and the cosmic microwave background itself is smooth
to one part in 100,000.

In the big bang cosmology of the 1960s and 1970s, the size and
smoothness of the observable universe were accepted as initial
conditions --- just the way it is.  In 1980, Alan Guth proposed
the theory of cosmic inflation, which {\it explains} this size
and smoothness as a consequence of exponential expansion 
during the first billionth of a billionth of a trillionth of
a second of cosmic history.  This ultra-fast, accelerating
expansion ironed out any wrinkles present in the pre-inflationary universe.
At the end of the inflation, the universe heated to high
temperature and thereafter followed the track of the standard
hot big bang theory.  Quantum noise --- the inevitable consequence of 
the famous Heisenberg Uncertainty Principle --- was stretched
from subatomic distances to macroscopic scales during inflation, 
providing the seeds of cosmic structure that would grow to
fruition billions of years later.

Guth's proposal was quickly adapted and refined by other
physicists including Andreas Albrecht, Richard Gott,
Stephen Hawking, Andrei Linde, Alexei Starobinsky, and 
Paul Steinhardt.  The most exotic of these refinements
is the ``eternal inflation'' scenario proposed by Linde in 1982,
according to which our entire observable universe is but one
of many (perhaps infinitely many) ``bubbles'' in an inflationary
sea, whose continuing exponential expansion pulls bubbles away 
from each other so rapidly that they have no chance to collide.
What appears to us as the ``big bang'' origin of the cosmos is
really just the moment when our bubble separated out from the
inflationary background, heated up, and and switched to 
decelerating expansion.

The theory of inflation has been highly successful, providing 
an explanation for the size and smoothness of the universe
and accurate quantitative
predictions for the wide variety of structures that
it contains.
The ``multiverse'' implied by Linde's eternal inflation model
is far more speculative, and the existence
of many disconnected cosmic bubbles may not be testable even in 
principle.  The multiverse scenario becomes still more exotic
when one allows the possibility that each bubble has
distinct physical properties, because of different durations
of inflation, different amounts of dark matter and dark energy,
perhaps even different fundamental laws governing atoms and
elementary particles, or different numbers of spatial dimensions.

It is this big, multi-faceted cosmos that {\it Island Universe}
depicts.  Each of the five sculptures represents its own bubble
universe, with the same conceptual structure and visual language
as {\it An End to Modernity}: an opaque last scattering surface,
quasar lamps, galactic clusters composed of glass disks and spheres,
and time increasing from the center to the outside.  We should imagine that 
each of these bubbles will continue to expand and grow new
structures at its future edge, but that the room in which we
view them is filled with energy whose repulsive gravity accelerates
the bubbles away from each other, so that their expanding
surfaces can never meet.  The lowest of the spheres began
growing earliest and has reached the largest size, while
the uppermost sculpture is small because it has had the least time
to grow.

The history and form of structure in each island universe is distinct,
corresponding to varying levels of quantum noise at the end of inflation or 
to varying amounts of gravitationally attractive dark matter
and gravitationally repulsive dark energy.  
After two days of discussions at the Institute for Advanced Study,
in fall 2006, Josiah and I had covered a wall's worth of blackboards
with sketches and notes about the five cosmic bubbles we had
chosen to represent.  We gave each universe a working title,
and these initial names, despite imperfections, have stuck 
throughout the project.
The central sculpture, ``Heliocentric,'' follows the same rules as
{\it An End to Modernity,} adjusted for its slightly smaller
size.  This sphere represents our own cosmic island, and the
Milky Way could be any one of the glass disks at its surface.
Or perhaps it is another bubble universe, much like our own
but unimaginably distant, and forever beyond our reach.

In ``Small Scale Violence,''
the largest and lowest sphere, inflation
has produced intense ripples on small scales, seeds that grow
quickly to produce the first galaxies soon after the epoch
of last scattering.  These galaxies are strongly clustered, and
frequent gravitational collisions of disks lead to a large
population of elliptical galaxies.  
Compared to ``Heliocentric,'' 
``Small Scale Violence'' has a high proportion of short rods
that contain these early forming galaxies, and, uniquely
among the five {\it Island Universe} sculptures, it contains
more spheres than disks.
Violent mergers also feed
the rapid and sustained growth of supermassive black holes.
This universe is bright with quasars, 
represented by a large number of high-intensity lamps.
From its inner epochs to its outer edge, ``Small Scale Violence''
is rich with large scale structure, containing the most
complex and populous clusters of {\it Island Universe}.

In ``Frozen Structure,'' a high proportion of dark energy 
truncates the growth and clustering of galaxies soon
after it has begun.
The first galaxies again form soon after recombination, 
clusters of small disks on short rods.
Once dark energy starts to drive accelerated cosmic expansion,
the gravitational attraction of dark matter 
cannot pull fresh material into the galaxies or pull the
galaxies together into larger structures.
``Frozen Structure'' does contain a handful of long rods, but the
galaxies and clusters that they hold are like those of the shorter,
earlier rods.
With few disordering collisions, this universe is dominated
by disk galaxies. 
After a brief quasar era, its undernourished
black holes fade away, and only the inner zones of
``Frozen Structure'' have illuminating lamps. 
In a universe dominated by dark energy, galaxies speed away from
each other to evolve in quiet isolation.

As viewed from Earth, the stars of our own Galactic disk
form a diffuse band of light, like a trail of milk spilled
across the sky.  Only when large telescopes allowed us to map
the distribution of distant galaxies did we learn that our cosmos
becomes uniform on large scales.  
In ``Directional Structure,'' the fourth island, an incomplete
end to inflation has left the universe anisotropic on the
largest scales, and rods emerge only along a tilted equatorial
belt.  Galaxies and quasars form only within this
thick, disk-shaped zone, and even this is unbalanced, with the
largest clusters all on one side.  As astronomers in this
universe trained their telescopes beyond the nearby stars,
they would find that distant galaxies were themselves arrayed
in a cosmic scale Milky Way, and that peering perpendicular
to this ``Super-Galaxy'' would reveal only formless darkness.

``Late Emergence,'' the highest and smallest sculpture, has the lowest
amplitude of inflationary ripples at recombination --- a last
scattering surface that is ten times smoother even than our own.
This universe has a prolonged dark age, billions of years with
no stars or galaxies to illuminate it, so even the shortest of
its rods extend nearly to the ``present day.''
The spherical arrangement of these rods reflects the large
scale uniformity of this cosmos.
Eventually, gravity
can grow structure even from small seeds, and this universe
may have a rich future ahead.  At the moment captured by
{\it Island Universe}, it is bursting into bloom for the
first time.

In 1905, Albert Einstein made the startling discovery that
time is relative, its flow depending on the state of
motion of the timekeeper.  In a homogeneous universe, there
is a natural choice of timekeepers, those ``go-with-the-flow''
observers (like ourselves) who follow the cosmic expansion and
see an isotropic cosmos.  We can therefore speak unambiguously
about the passage of time and the history of structure within
any one cosmic bubble.  In eternal inflation, however, the
universe becomes radically inhomogeneous on scales far beyond
our cosmic horizon, and there is no one privileged set of
observers to run the clocks.  In {\it Island Universe}, therefore,
there is no clear way to reconcile the time of an individual
bubble, which increases radially outward from its center,
with the time of another bubble, or with the ``global'' time
that increases from the floor to the ceiling.  As we walk
among the islands, wandering superluminally through an inflationary
sea whose smooth expanse and bubbling froth must extend far
beyond the gallery walls, we assume vantage points that can
exist only in the abstract realms of science, mathematics,
and art.

\bigskip
\noindent
{\bf Annotated Bibliography}

\parindent=0pt

{\it Josiah McElheny: A Prism}, ed. L. Neri \& Josiah McElheny,
Skira/Rizzoli books, New York, 2010.  The photographs and essays
in this book cover all of Josiah's work, including the cosmological
sculptures.  

{\it Josiah McElheny -- Island Universe}, ed. J. McElheny, White Cube, 
London, 2009.  This is the catalog from the London exhibition of
{\it Island Universe}, including many images, an edited ``conversation''
between Josiah and me about the project, a crucial chapter from
Kant's {\it Universal History and Theory of the Heavens}, a reprint
of Andrei Linde's article {\it Particle Physics and Inflationary Cosmology},
and essays by art historian Molly Nesbit, cosmologist Craig Hogan,
and philosopher Thomas Ryckman.

{\it A Space for an Island Universe}, ed. L. Cooke and J. McElheny,
Museo Nacional Centro de Arte Reina Sofia, Madrid, 2009.
This is the catalog from the Madrid exhibition of {\it Island Universe},
including much of the material in the London catalog plus images and
texts about the Madrid exhibition and about the companion film,
also titled {\it Island Universe}.

{\it Josiah McElheny: Notes for a Sculpture and a Film}, ed. J. McElheny
and H. Molesworth, Wexner Center for the Arts, Columbus, 2006.  
This is the catalog for the Wexner Center exhibition of 
{\it An End to Modernity} and includes images of and essays about
the sculpture and the accompanying film 
{\it Conceptual Drawings for a Chandelier (1965)}.

From the Big Bang to {\it Island Universe}: Anatomy of a Collaboration, 
by D. H. Weinberg, to appear in the journal {\it Narrative} as part
of the proceedings of {\it Narrative, Science, and Performance},
ed. J. Phelan.  This article describes the history of our collaboration.
Also available on {\tt arXiv}.

Links to higher resolution images and other information about
the sculptures can be found at
{\tt http://www.astronomy.ohio-state.edu/$\sim$dhw/McElheny}.

\vskip 1.0truein
\input epsf

\centerline{
\epsfxsize=6.40truein
\epsfbox{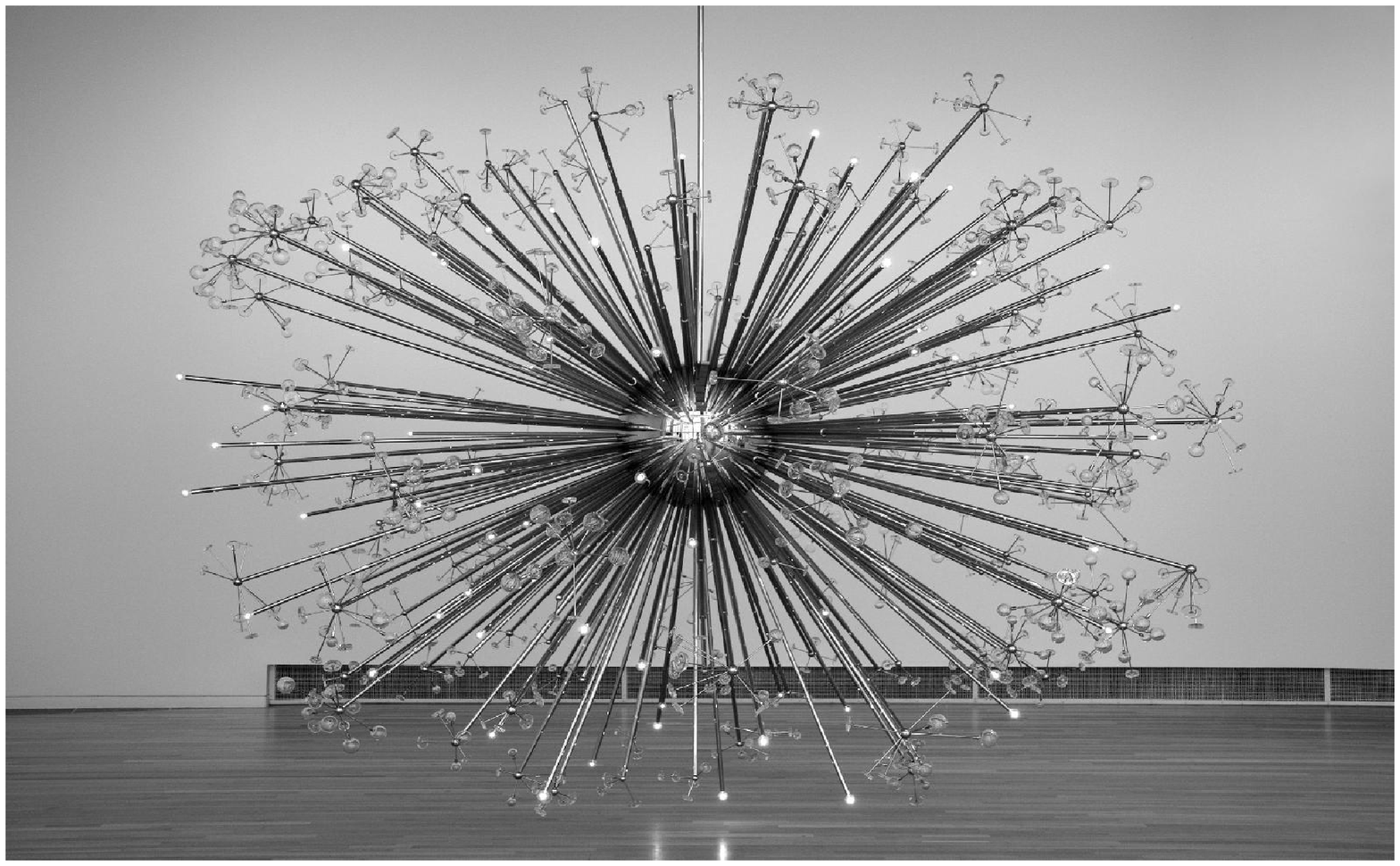}
}

\vskip 5pt

\centerline{
\epsfxsize=2.10truein
\epsfysize=1.40truein
\epsfbox{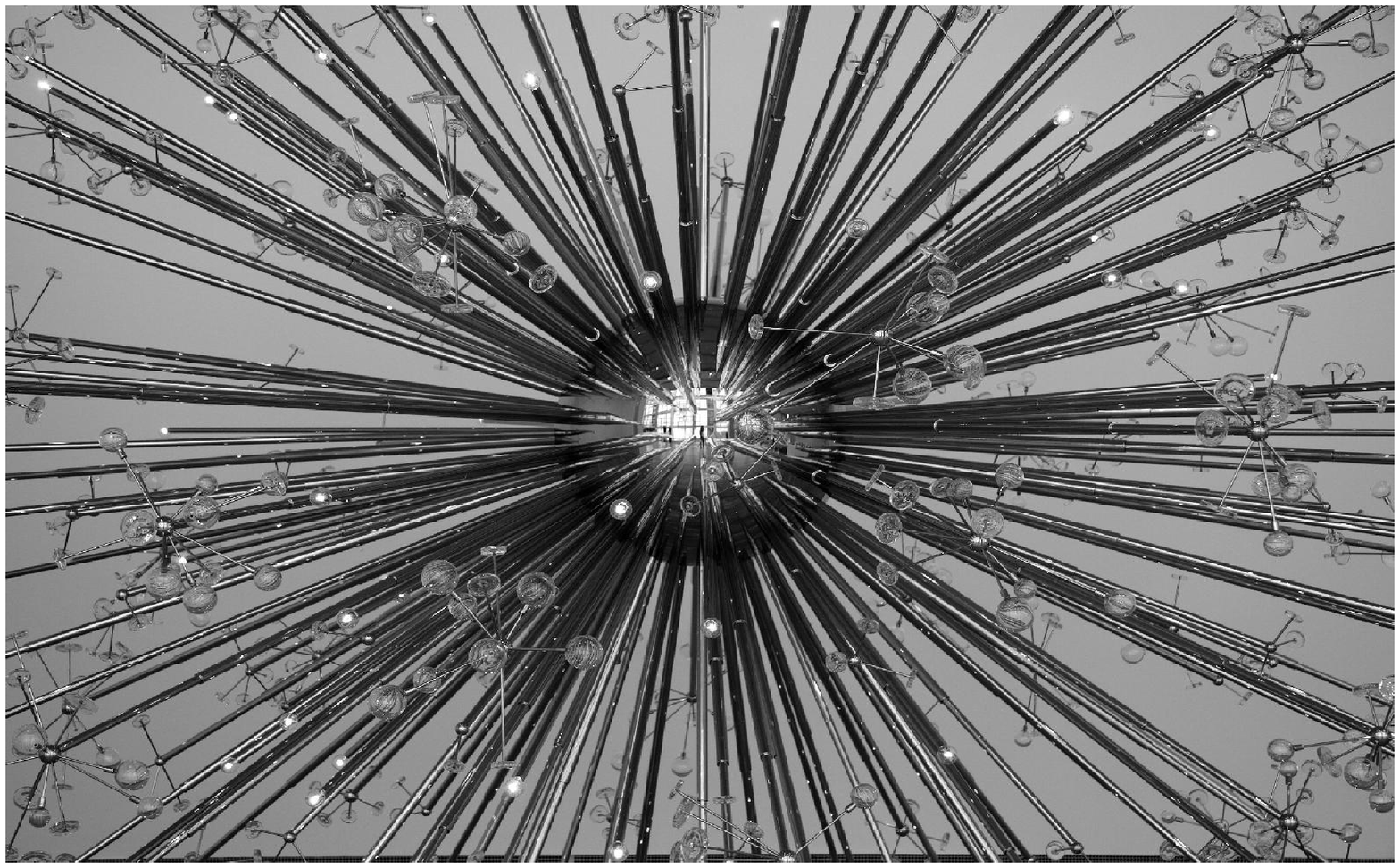}
\hfill
\epsfxsize=2.10truein
\epsfysize=1.40truein
\epsfbox{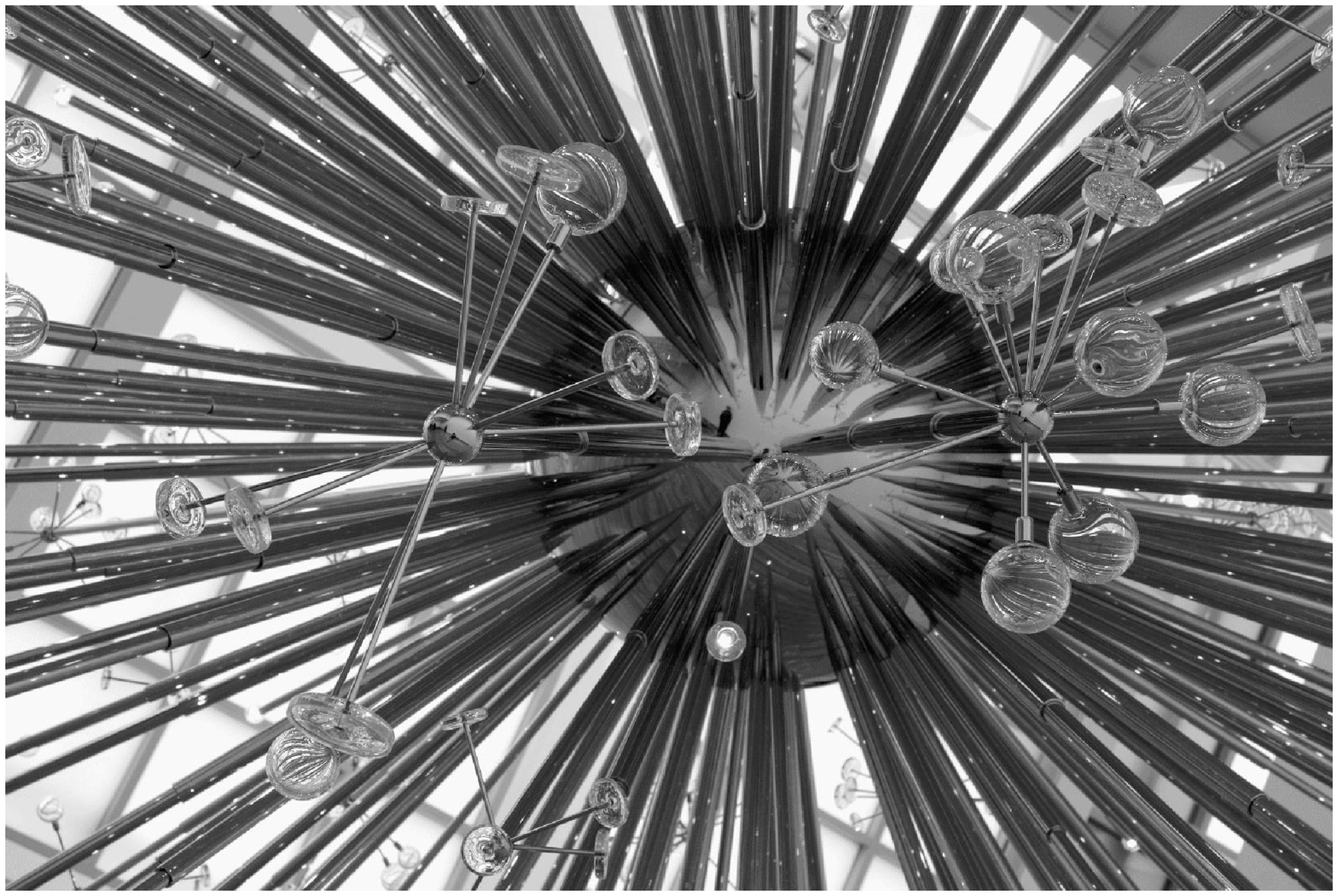}
\hfill
\epsfxsize=2.10truein
\epsfysize=1.40truein
\epsfbox{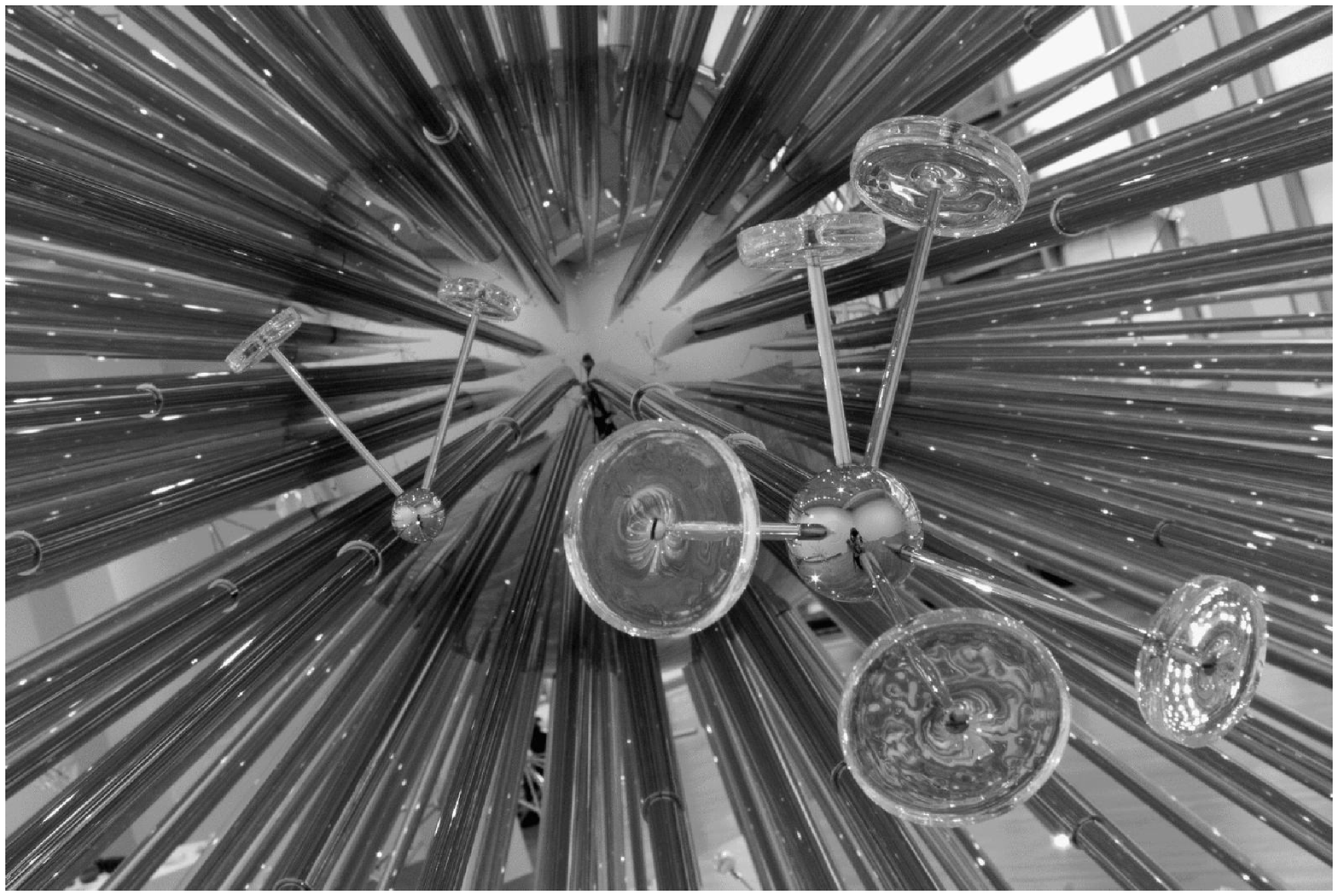}
}

\medskip
\noindent
Four views of {\it An End to Modernity} (Josiah McElheny, 2005),
exhibited at the Wexner Center for the Arts, Ohio State University, Columbus,
Ohio.  Dimensions are $172"\times 138"$.
{\it An End to Modernity} is now in the permanent collection of
the Tate Modern, London.  Image courtesy of J. McElheny.

\vfill\eject

\bigskip

\centerline{
\epsfysize=3.5truein
\epsfbox{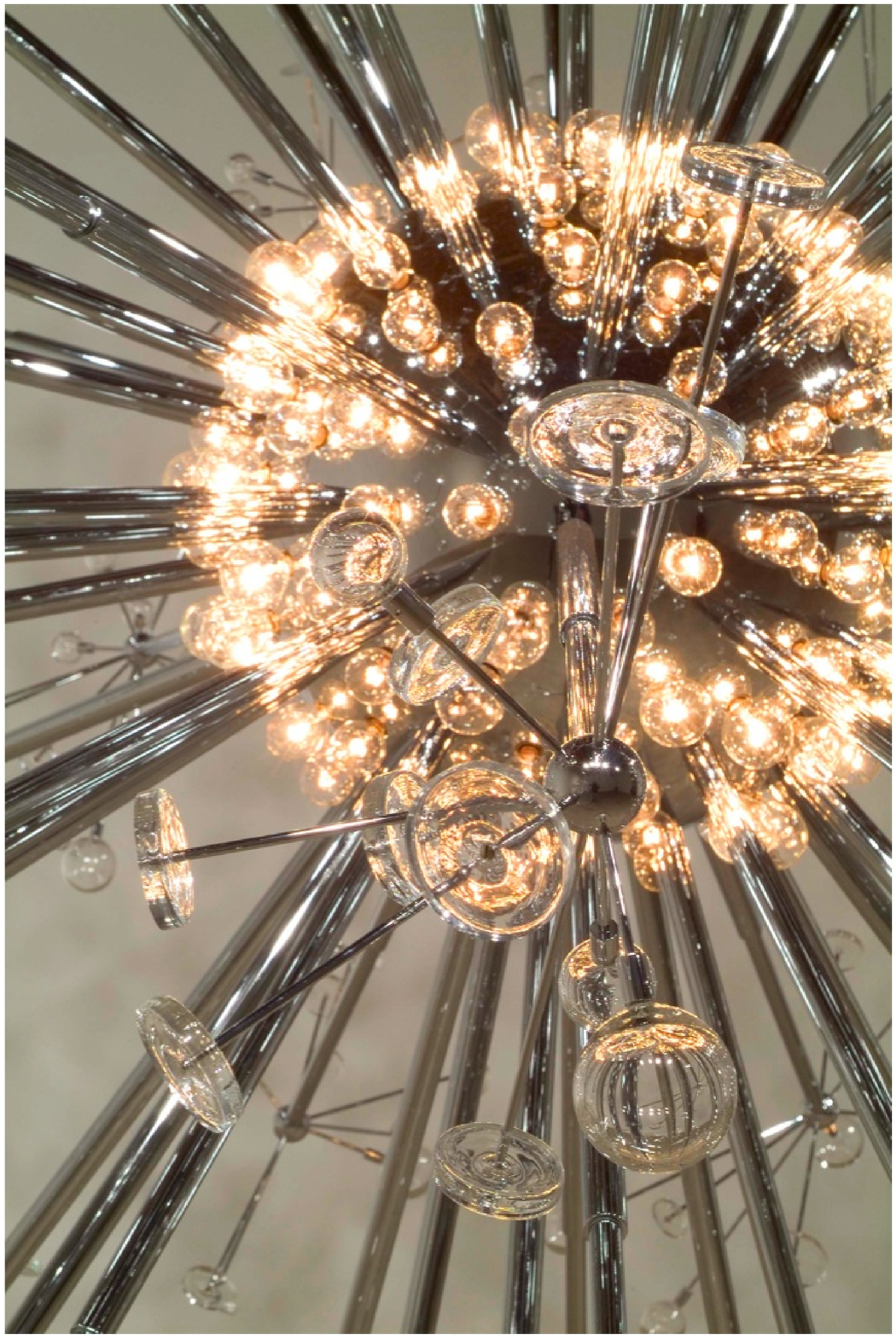}
\hfill
\epsfysize=3.5truein
\epsfbox{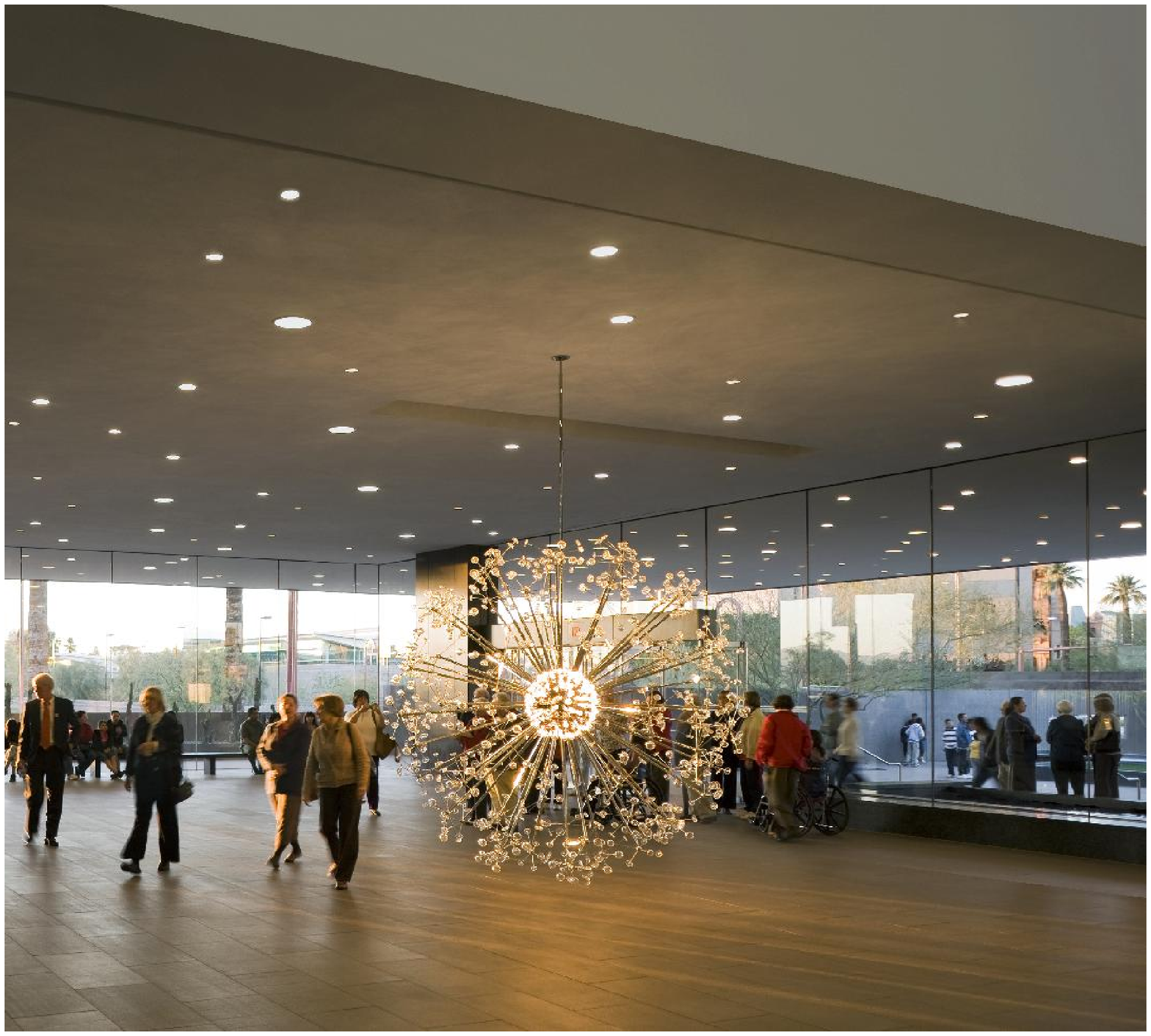}
}

\medskip
\noindent
{\it The Last Scattering Surface} (Josiah McElheny, 2006)
exhibited at the Donald Young Gallery in Chicago (left, detail) and
in its permanent installation at the Phoenix Art Museum (right).
Dimensions are $88"\times 88"$.
Images courtesy of J. McElheny.

\medskip

\centerline{
\epsfysize=3.5truein
\epsfbox{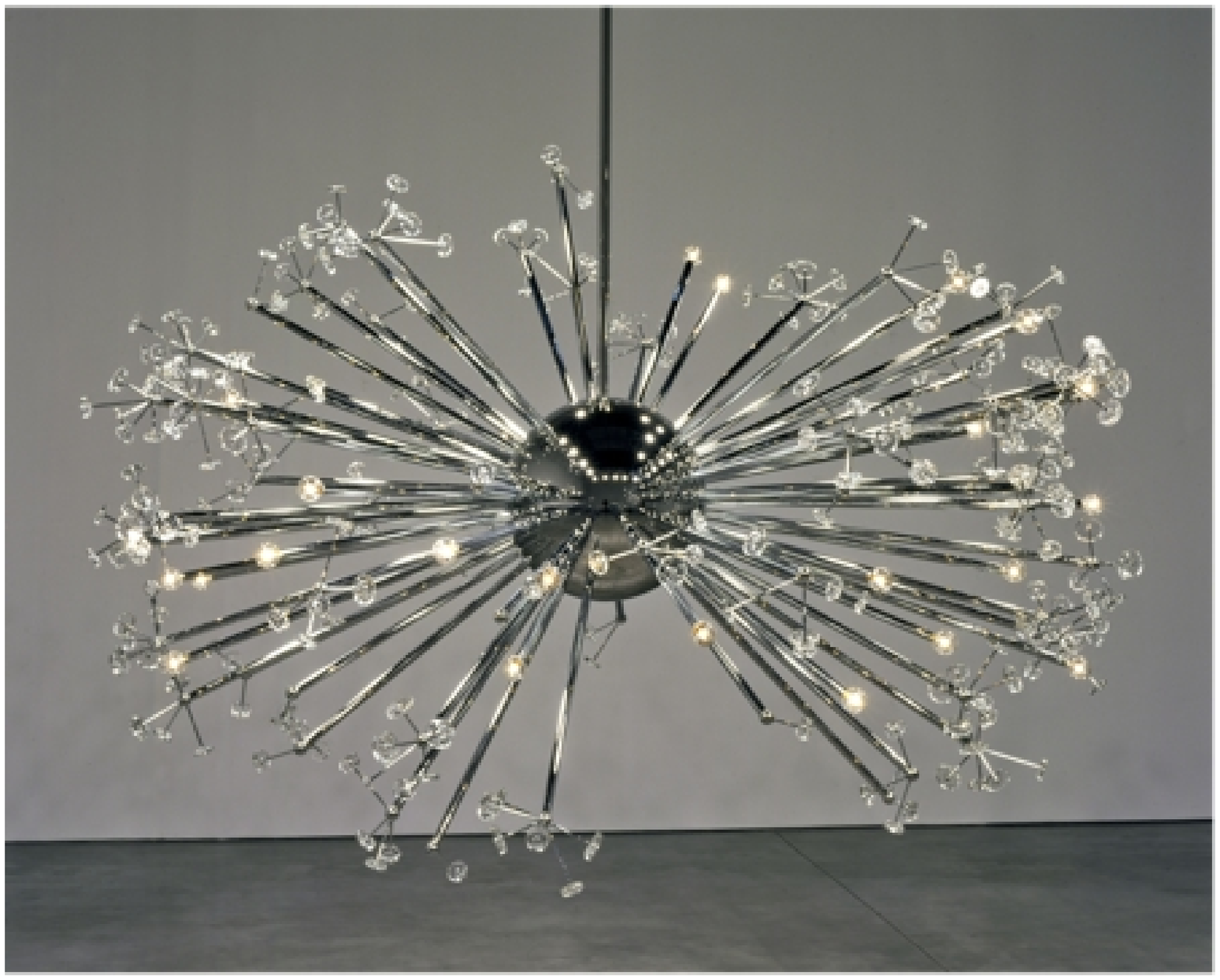}
}

\medskip
\noindent
{\it The End of the Dark Ages} (Josiah McElheny, 2008),
exhibited at the Andrea Rosen Gallery in New York City.
Dimensions are $88"\times 66"$.
{\it The End of the Dark Ages} is now in a private collection.
Image courtesty of J. McElheny.

\vfill\eject

\centerline{
\epsfxsize=6.0truein
\epsfbox{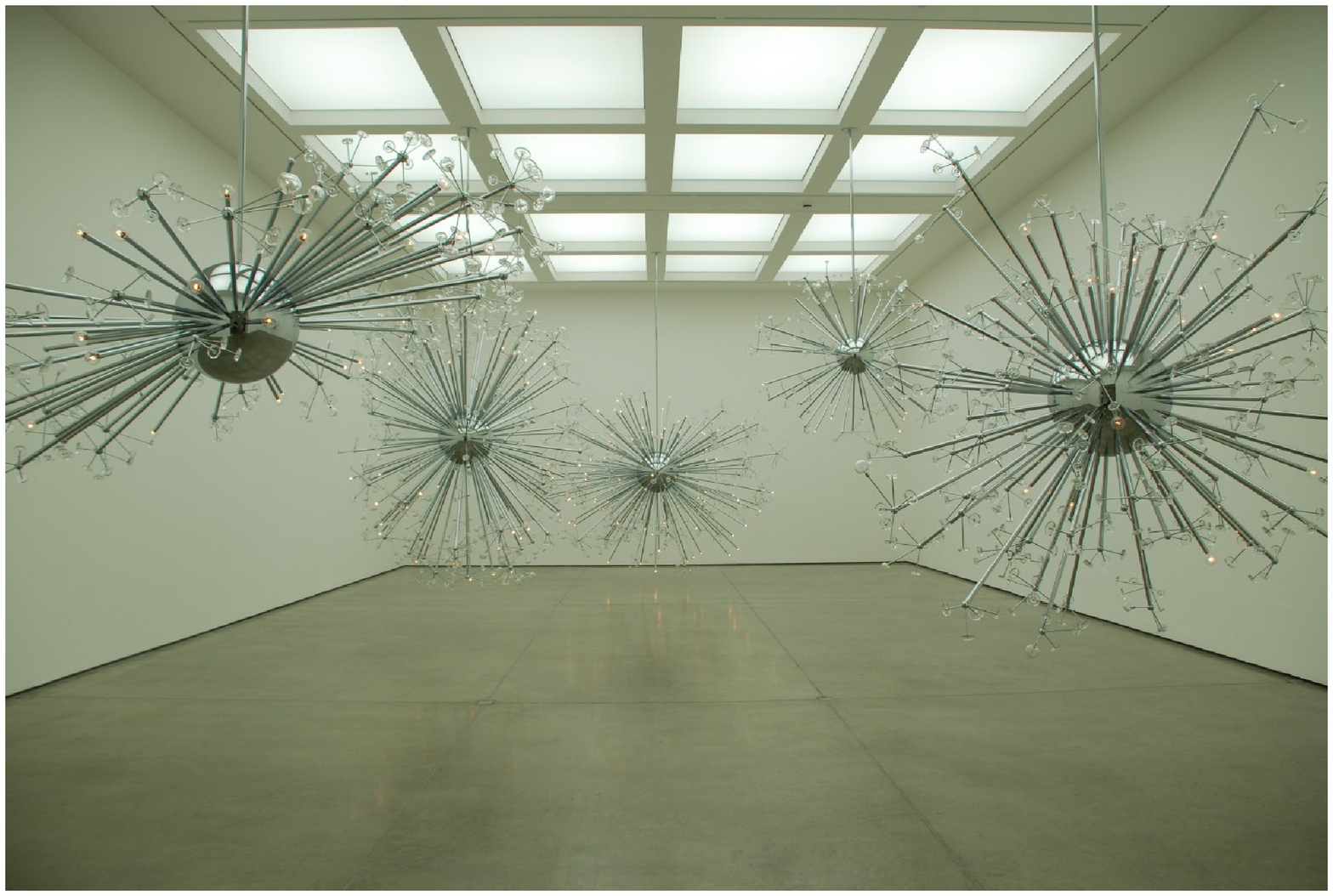}
}
\smallskip
\centerline{
\epsfxsize=6.0truein
\epsfbox{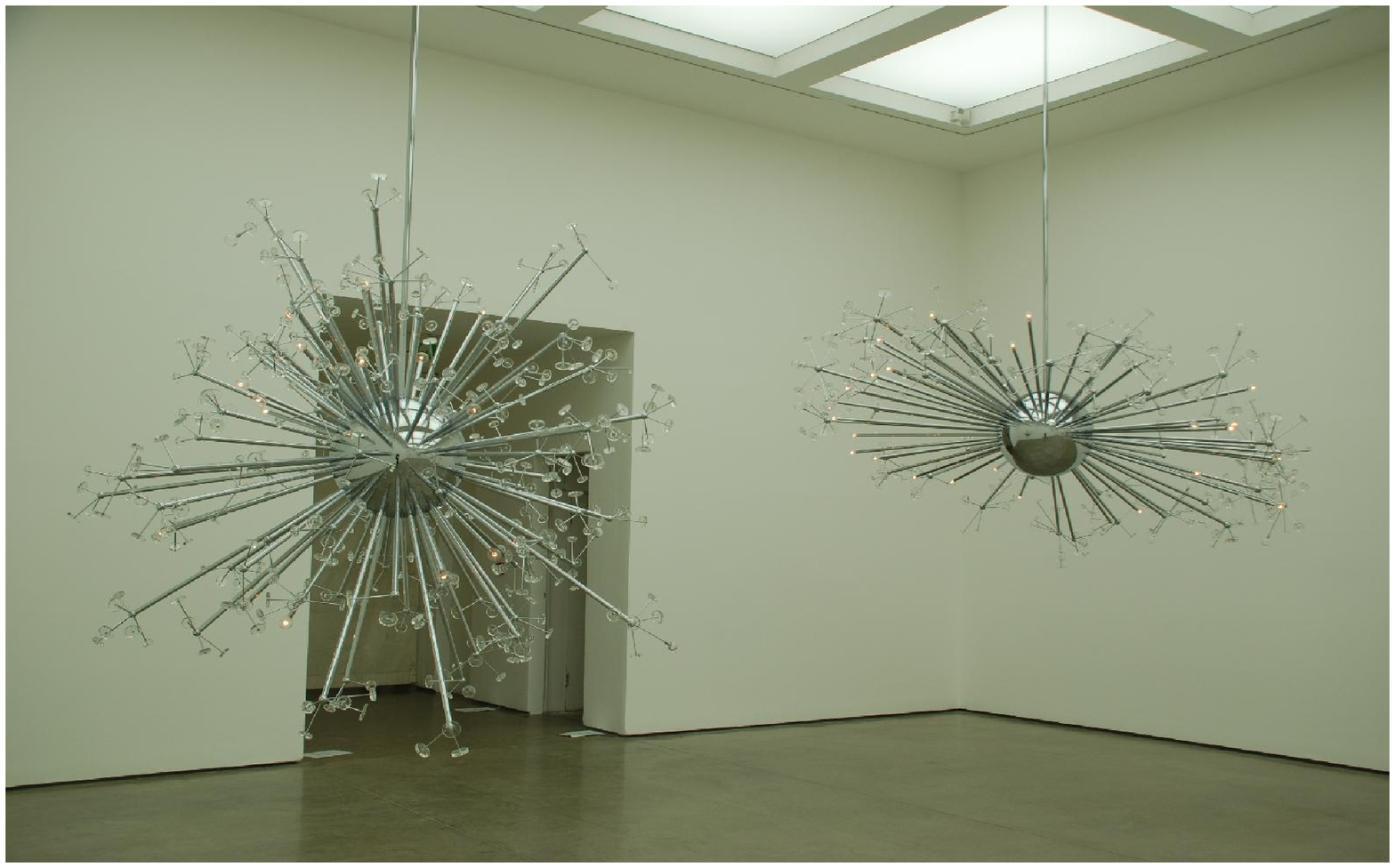}
}

\medskip
\noindent
{\it Island Universe} (Josiah McElheny, 2008), exhibited at the 
White Cube Gallery in London.  
Dimensions of individual elements range from $144"\times 116"$
to $74"\times 74"$.
The lower image shows the elements
``Frozen Structure'' and ``Directional Structure.''
Image courtesy of J. McElheny.

\bye